\documentclass[%
aps,
reprint,
superscriptaddress,
 amsmath,amssymb,
prc,
]{revtex4-1}

\usepackage[utf8]{inputenc}
\usepackage{gensymb}
\usepackage{graphicx}% Include figure files
\usepackage{dcolumn}% Align table columns on decimal point
\usepackage{bm}% bold math
\usepackage{mathtools}
\usepackage{epstopdf} % .eps figures
\usepackage{hyperref}
\usepackage{color}

\newcommand{\ket}[1]{| #1 \rangle}
\newcommand{\bra}[1]{\langle #1 |}
\newcommand{\expected}[1]{ \langle #1 \rangle}
\newcommand{\product}[2]{\langle #1 | #2 \rangle}

\begin{document}
\title{Vortices in low-density neutron matter and cold Fermi gases}

\author{Lucas Madeira}
\email{madeira@ifsc.usp.br}
\affiliation{Instituto de F\'isica de S\~ao Carlos, Universidade de S\~ao Paulo, CP 369, S\~ao Carlos, S\~ao Paulo 13560-970, Brazil}

\author{Stefano Gandolfi}
\affiliation{Theoretical Division, Los Alamos National Laboratory,
Los Alamos, New Mexico 87545, USA}

\author{Kevin E. Schmidt}
\affiliation{Department of Physics, Arizona State 
University, Tempe, Arizona 85287, USA}

\author{Vanderlei S. Bagnato}
\affiliation{Instituto de F\'isica de S\~ao Carlos, Universidade de S\~ao Paulo, CP 369, S\~ao Carlos, S\~ao Paulo 13560-970, Brazil}

\date{\today}

\begin{abstract}
Cold gas experiments can be tuned to achieve strongly-interacting
regimes such as that of low-density neutron matter found in neutron-stars' crusts.
We report $T$=0 diffusion Monte Carlo results (i) for the ground state
of both spin-1/2 fermions with short-range interactions
and low-density neutron matter in a 
cylindrical container, and (ii) properties
of these systems with a vortex line excitation.
We calculate the equation of state for cold atoms and low-density
neutron matter in the bulk systems, and we contrast it to
our results in the cylindrical container.
We compute the vortex line excitation energy for different
interaction strengths, and we find agreement between
cold gases and neutron matter
for very low densities.
We also
calculate density profiles, which allow us to determine the
density depletion at the vortex core, which
depends strongly on the short-ranged interaction in cold atomic gases, but it is of $\approx$ 25\% for neutron
matter in the density regimes studied in this work.
Our results can be used to constrain neutron matter properties
by using measurements from cold Fermi gases experiments.
\end{abstract}

\maketitle

\section{Introduction}

Strongly interacting fermionic systems appear in many contexts, for example: 
superconductors, cold atomic Fermi gases, low-density neutron matter, and
QCD at high baryon densities.
Shedding light on properties of one of these systems
may contribute to our comprehension of 
strongly interacting
Fermi systems as a whole.

Cold atom systems provide an example where the interplay between experiments
and theory led to rapid advances in the field.
In these dilute systems, short-range interactions are characterized by a single parameter $k_F a$, the
product of the Fermi wave number $k_F$ and the
$s$-wave scattering length $a$. This interaction strength can be tuned
using an external magnetic field near a Feshbach resonance, and
the attractive interactions
can span
a continuum between
the Bardeen-Cooper-Schrieffer (BCS) limit of superfluidity and the Bose-Einstein condensation (BEC) of dimers, passing through the unitary limit of
infinite scattering length.
Experiments with cold atoms can provide direct tests of quantities such
as the equation of state and pairing gap, which are currently
inaccessible to their neutron matter counterparts.
For a review on the subject the reader is referred to Ref.~\cite{gio08} and
references therein.

On the other hand,
if we compare cold gases to neutron matter, we find that
the neutron-neutron interaction can be more complicated: short-range
repulsion, two-pion exchange at a intermediate range, and one-pion
exchange at large distances. However, this situation changes in the
low-density regime, which is the case in the exterior of neutron-rich
nuclei and neutron-star crusts. In these systems, the scattering length
and effective range of the interaction
are the most essential quantities for describing the
physical properties, and properties of neutron matter and cold
atoms are similar~\cite{Carlson:2012,Gandolfi:2015}.
%Higher density states are not energetically favored
%due to the short-range repulsive core of the neutron-neutron interaction.

A neutron matter model with a zero-range interaction \cite{Baker2001} was presented as a ``many-body challenge" proposed by Bertsch,
\footnote{ The challenge proposed to the participants of the Tenth International Conference on Recent Progress in Many-Body Theories
can be stated as:
what are the ground state properties of the many-body system composed
of spin-1/2 fermions interacting via a zero-range, infinite scattering
length contact interaction?
}, much before cold atom experiments could shed light on the properties of these systems.
In dilute cold gases, the effective range $r_e$ between atoms is
much smaller than the interatomic spacing $r_0$, and can be taken to be
zero. The diluteness can guarantee
that the scattering length $a$ is much larger than $r_0$. Comparison
with other systems is meaningful if they also obey 
$|a| \gg r_0 \gg r_e$.
The scattering length of neutron matter, $a^{nn} \sim -18.5$ fm, is
substantially larger than the interparticle distance and the effective range,
$r_e^{nn}\sim 2.7$ fm, such that $|r_e^{nn}/a^{nn}|\approx 0.15$.
However, only at very low-densities is the effective range much smaller
than the interparticle distance.  If we neglect the effects of a finite
effective range in the neutron-neutron interaction, cold atoms and neutron
matter are universal in the sense that properties depend only on the product
$k_F a$.

Quantum Monte Carlo (QMC) methods have been successful at comparing
the equation of state and pairing gap of cold atom systems and
low-density neutron matter \cite{Gezerlis:2008,gez10}. In the
present
work we used a similar model to compare properties of vortices
in low-density neutron matter and cold Fermi gases.
One signature of superfluidity is the formation of quantized vortices,
where
the quantization of the flow is given in units of $h/(2m)$, $m$
being the mass of the fermion.
The
microscopic structure of a vortex line in neutron matter
has been studied
using Bogoliubov-de Gennes equations \cite{bla99,Elgaroy2001}, and nuclear energy density functional approaches \cite{yu03}.
For cold atom gases there is an abundance of studies,
for example $T=0$ results using
Bogoliubov-de Gennes equations at unitarity \cite{bul03} and
throughout the BEC-BCS crossover \cite{sen06}, and
finite temperature calculations \cite{sim13}.

Here we report results for a single vortex line in a cylindrical 
geometry
for both low-density neutron matter and cold Fermi gases using QMC
methods.
We investigated the consequences of the finite effective range of the
neutron-neutron interaction, in contrast to $r_e \approx 0$ for cold gases.
We also studied effects that go beyond low-energy scattering
by using two potential models for the neutron-neutron interaction,
one based on phenomenology, and another that was tuned to reproduce
the desired low-energy phase shifts.
We calculated the equation of state for cold atoms and low-density
neutron matter in the bulk systems.
We show that it is possible to separate the energy contributions
of systems in a cylindrical container with hard walls into bulk
and surface terms.
The excitation energy necessary to produce a vortex line
was computed by
using the energy difference between a system of pairs with
angular momentum $\hbar$ and the ground state.
We show that for very low densities there is an agreement between
the excitation energies for vortex-line formation between cold gases
and neutron matter. However, as the density increases (or as the interaction
strength increases in absolute value) they differ.
We also
calculated density profiles, which allows us to determine the
density depletion at the vortex core. We found that the depletion varies
from 28\% up to 47\% for cold gases, whereas for neutron matter the depletion
is approximately 25\%, for the density range we studied in this work.
Our results are compared
to previous mean-field calculations.

This paper is structured as follows.
In Sec.~\ref{sec:met} we introduce our methodology.
We discuss aspects of the
cylindrical container in Sec.~\ref{sec:cyl}, and
low-energy two-body scattering in
Sec.~\ref{sec:scatt}.
We present the wave functions we built in
Sec.~\ref{sec:wf}, which describe properties of the bulk systems,
and systems in a cylindrical container (both the ground state and
systems with a vortex line).
In Sec.~\ref{sec:qmc} we give a brief description of the QMC methods we
employed.
Section \ref{sec:res} presents our results,
namely the ground state and vortex excitation energies in
Sec.~\ref{sec:energy}, and density profiles in Sec.~\ref{sec:dens}.
An outlook is provided in Sec.~\ref{sec:sum}.
Finally, in Appendix~\ref{app} we show how to obtain an exact relationship between
scattering length and the parameters of the modified Poschl-Teller potential.

\section{Methods}
\label{sec:met}

\subsection{Cylindrical container}
\label{sec:cyl}

The choice of which trapping potential (or geometry) to use in this problem
is not unambiguous, as there is a trade-off for each possible candidate.
A choice that minimizes surface effects is to have
an array of counter rotating vortices with
periodic boundary conditions. One drawback is that this state has zero
total angular momentum, thus it can decay to the ground state of the system.
Also, from the computational perspective, this choice is not feasible for fermionic
systems. For example, $^4$He
calculations of Ref.~\cite{sad97} used 300 particles and four counter
rotating vortices in the simulation cell. In order to use the same
number of fermion pairs we would require a system of 600 fermions.
Another possible choice would involve
harmonic traps, which are readily available in experimental setups, however,
the density profiles of cold gases in harmonic traps can also
differ substantially from what is expected in the thermodynamic limit 
\cite{cha07}.

Instead, we opted for using a cylindrical container
of radius ${\cal R}$ and height ${\cal L}$,
with hard walls, periodic 
in the axial direction. This choice is consistent with previous bosonic \cite{vit96}
and fermionic \cite{mad16} calculations.
Also, this is the generalization of the two-dimensional (2D) disk geometry to 3D
\cite{Ortiz:1995,gio96,mad17}, where we made the axial direction periodic.
Throughout this work we use $(\rho,\varphi,z)$ to denote the
usual cylindrical coordinates.

In the thermodynamic limit, ${\cal R,L}\to\infty$, the energy per particle is independent
of the cylinder radius and height, and it should
go to the bulk value.
The relationship between thermodynamic properties of a confined fluid
and the shape of the container is often expressed as a function of
the various curvatures of the container \cite{kon04}. For these reasons,
we chose the following functional form for the energy per particle
in the cylindrical geometry:
\begin{eqnarray}
\label{eq:ERL}
E^{\rm cyl}({\cal R},{\cal L})=E_0^{\rm cyl}+\frac{\lambda_S}{2\pi{\cal R}{\cal L}},
\end{eqnarray}
where $E_0^{\rm cyl}$ represents a bulk contribution to the energy, and the
second term on the right-hand side is a ``surface'' contribution.
Corrections to this functional form would come in powers of ${\cal R}^{-1}$
and/or ${\cal L}^{-1}$, however, we found those do not improve the description
of the results.

One of the complications of introducing hard walls is the presence of the
so-called Friedel oscillations.
Fermionic systems bound by hard walls display density profiles 
characterized by Friedel oscillations. Although they are present in
three dimensions, they are
more pronounced in low-dimension systems such as 1D \cite{mck16}, and 2D \cite{mad17}.
We would like our system to exhibit some desirable features with
respect to the energy and density distribution ${\cal D}(\rho)$. 
Regarding the density distribution as a function of the radial 
coordinate (see Sec.~\ref{sec:dens} for the
normalization and profiles in the interacting cases),
besides a vanishing density at the walls, 
we want the profile to be flat close to the axis of the cylinder.
This would be the behavior in the thermodynamic limit, but this is not always true for finite-size systems.
If we fix the number density at $n=k_F^3/(3\pi^2)$, the free Fermi gas density, we have freedom to choose either the cylinder radius ${\cal R}$ or the height ${\cal L}$.
In making this choice we adopted the following procedure. We calculated analytically the energy and density profile for the free gas, and we looked for systems that obeyed the criteria established above, that is:
(i) the energy of the system for different particle numbers $N$ is well described by Eq.~(\ref{eq:ERL});
(ii) the slope of the density profile in the vicinity of the origin
($\rho \leqslant \rho_0$)
is less than a prescribed tolerance, $|\partial {\cal D}(\rho)/\partial \rho|_{\rho\leqslant\rho_0} \leqslant \epsilon$;
(iii) density oscillations are minimized.

\begin{figure}[!htb]
\centering
\includegraphics[angle=-90,width=\linewidth]{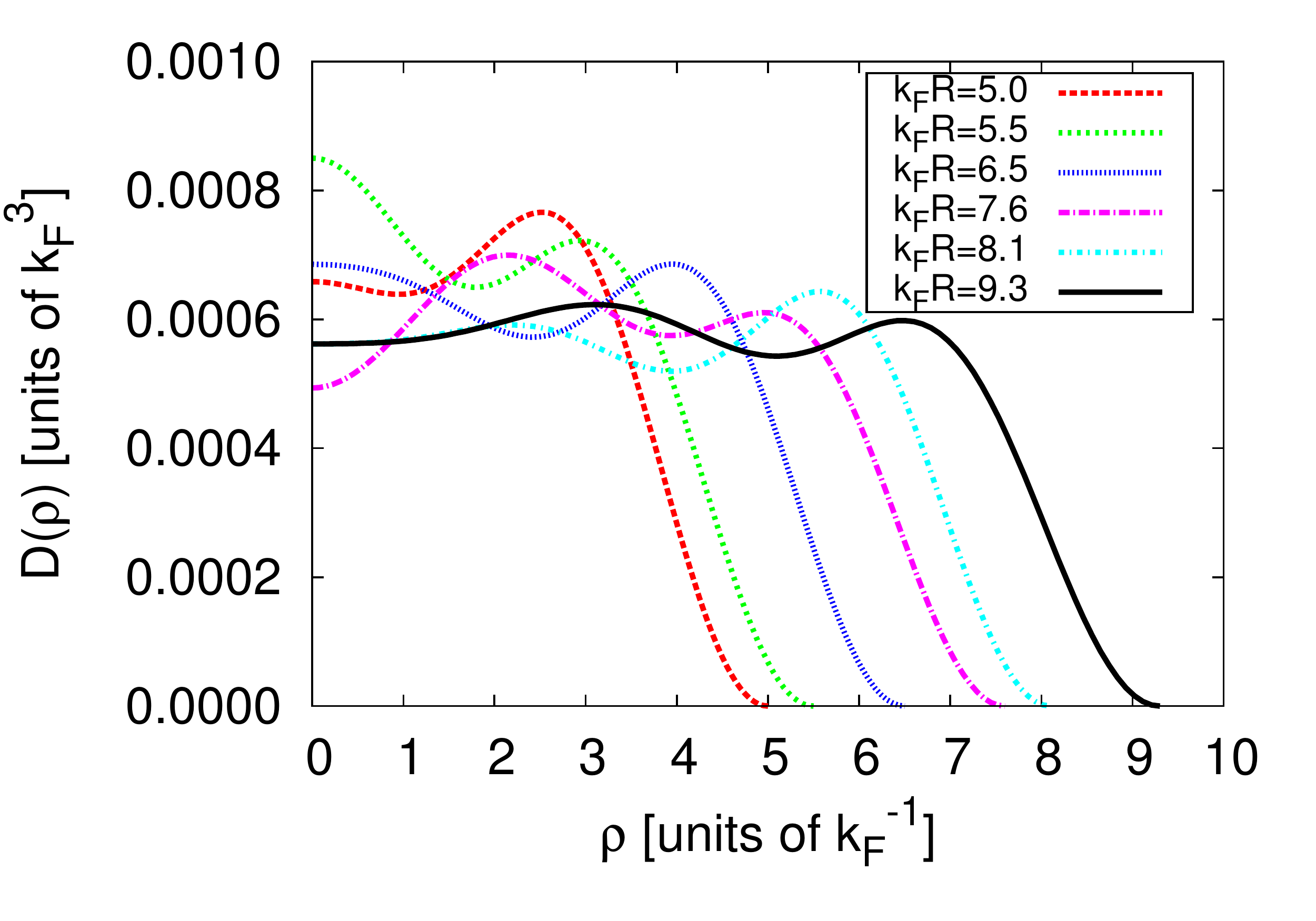}
\caption{(Color online) Density profile of the free
Fermi gas as a function
of the radial coordinate $\rho$ for $N=78$ and several
radii:
5.0$k_F^{-1}$ long dashed (red) line,
5.5$k_F^{-1}$ dashed (green) line,
6.5$k_F^{-1}$ short dashed (blue) line,
7.6$k_F^{-1}$ long dashed-dotted (magenta) line,
8.1$k_F^{-1}$ dashed-dotted (cyan) line, and
9.3$k_F^{-1}$ continuous (black) line.
The number density was kept fixed at the free Fermi gas value,
$k_F^3/(3\pi^2)$.
Although the behavior close to $\rho={\cal R}$ is similar
for all radii, due to the presence of the hard walls,
the profile at the center of the cylinder,
$\rho \lesssim 2.0k_F^{-1}$, can be quite different.
In our simulations we employed ${\cal R}=9.3k_F^{-1}$ for $N=78$.
}
\label{fig:dens_cyl_free}
\end{figure}

Throughout this work we report energies per particle in units of the free Fermi gas energy per particle,
\begin{equation}
\label{eq:EFG}
E_{FG}=\frac{3}{10}\frac{\hbar^2}{m}\left(3\pi^2\frac{N}{V}\right)^{2/3}.
\end{equation}
We found that for $N=\{78,80,82,84,86\}$, with the radius
${\cal R}=9.3k_F^{-1}$ for $N$=78 and $9.4k_F^{-1}$ for
the other systems, are well described by Eq.~(\ref{eq:ERL}) with $E_0^{\rm cyl}=1.02(2)E_{FG}$ and $\lambda_S=108(8)E_{FG}k_F^{-2}$,
and the maximum value of $\epsilon$ is of $\sim 5\times10^{-5} k_F^4$ for $\rho_0\leqslant 2 k_F^{-1}$.
To illustrate how criteria (ii) and (iii) are not easily met, we plot in Fig.~\ref{fig:dens_cyl_free} the density profile for the free 
Fermi
gas with $N=78$ and several different values of ${\cal R}$. 
Our ansatz takes into consideration only the free gas case, which 
corresponds to the $-k_Fa\to 0$ limit. However, we show in Secs.~\ref{sec:gscyl} and \ref{sec:dens} that 
our choices produced the desired results in the
$0.5\leqslant -k_F a \leqslant 5.0$ range.

\subsection{Scattering}
\label{sec:scatt}

Two-body scattering for a finite range potential $V(r)$ is described by the Schr\"odinger equation.
We separate the solutions into radial and angular parts, with the latter being a constant for $s$-wave scattering.
The scattering length $a$ and the effective range $r_e$ can be determined from the zero-energy solution of the radial equation and its asymptotic form.
The low-energy behavior of the phase shift $\delta(k)$ can be related to $a$ and $r_e$ \cite{bet49},
\begin{eqnarray}
\label{eq:phase_shifts}
k \cot \delta(k)=-\frac{1}{a}+\frac{r_e k^2}{2}+\mathcal{O}(k^4),
\end{eqnarray}
hence different potentials that reproduce the same scattering length and effective range yield the same low-energy phase-shift behavior.
When simulating cold gases,
we chose the modified Poschl-Teller (mPT) potential to describe
interactions between antiparallel spins,
\begin{equation}
\label{eq:mpt}
V_{\rm mPT}(r)=-v_0\frac{\hbar^2}{m_r} \frac{\mu^2}{\cosh^2(\mu r)},
\end{equation}
where $v_0$ and $\mu$ are parameters that can be tuned to reproduce the desired $a$ and $r_e$.
We restricted the parameters so that no bound state is supported.
The quantities $a$, $\mu$, and $v_0$ are related through (see Appendix~\ref{app})
\begin{equation}
\label{eq:analytic}
a\mu = \frac{\pi}{2} \cot\left( \frac{\pi \lambda}{2} \right)+\gamma+\Psi(\lambda),
\end{equation}
where $\gamma=0.577\dots$ is the  Euler-Mascheroni constant, $\Psi$ is the digamma function, and $\lambda$ is such that $v_0=\lambda(\lambda-1)/2$.
In the equation above, the requirement on the number of bound states, and a fixed $r_e$, completely determine the parameters of the potential for a given scattering length.

\begin{figure}[!htb]
\centering
\includegraphics[angle=-90,width=\linewidth]{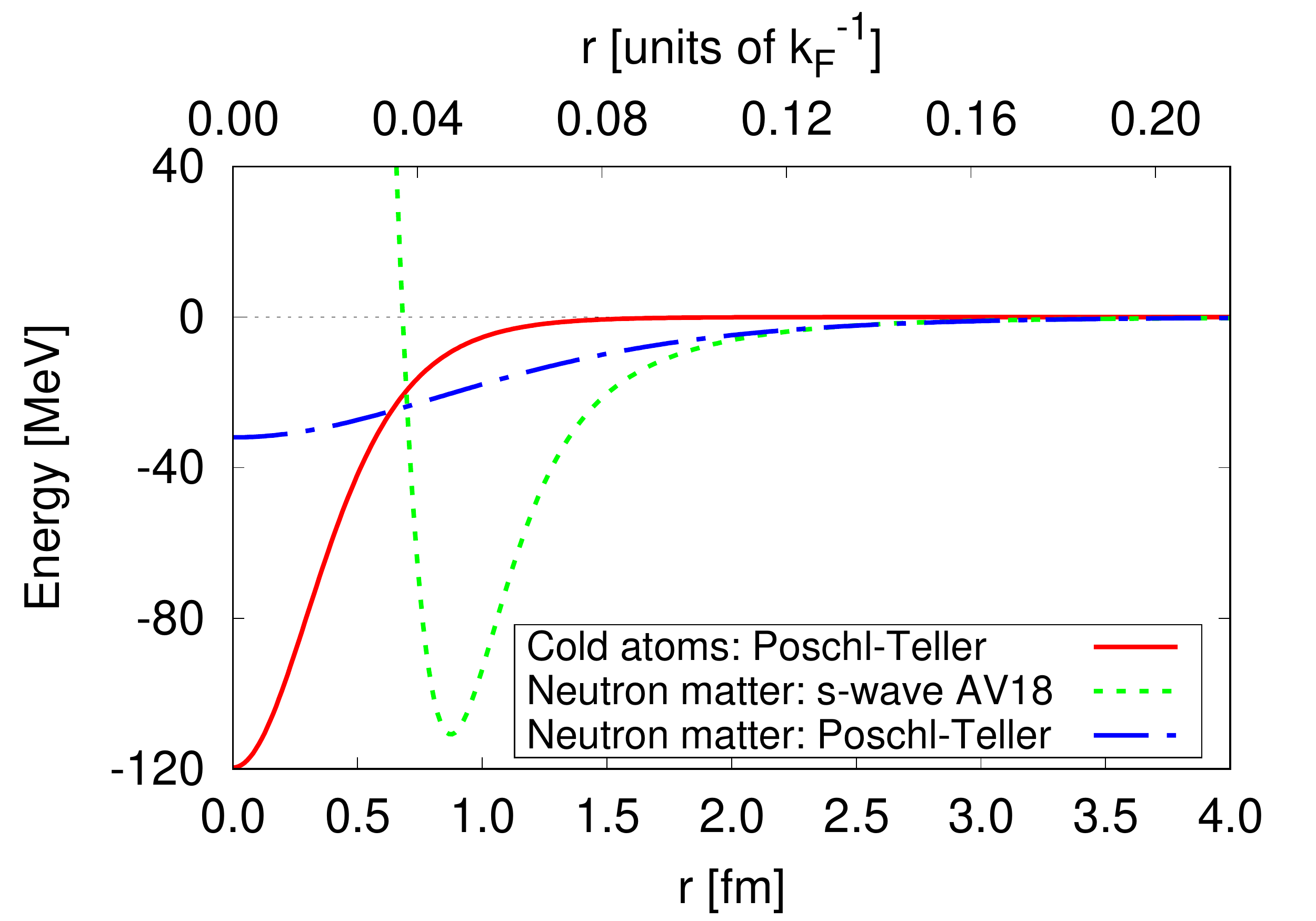}
\caption{(Color online)
Comparison between the pair-wise interactions employed in this work for $-k_Fa=1$.
The continuous (red) line denotes the modified Poschl-Teller potential,
Eq.~(\ref{eq:mpt}),
with $k_F r_e=0.05$,
the dashed (green) line the $s$-wave component of AV18,
and the dotted-dashed (blue) line stands for the
modified Poschl-Teller potential
tuned to reproduce
$a^{nn}=-18.5$ fm and $r_e^{nn}=2.7$ fm. The top $x$ axis
displays the distances in units of $k_F^{-1}$, considering the number
density to be the same as the free Fermi gas,
$n=k_F^3/(3\pi^2)$.
}
\label{fig:pot}
\end{figure}

For the neutron matter simulations, we employed two different potential
interactions. Our goal with this approach is to see if
there are any relevant effects beyond the low-energy regime described by Eq.~(\ref{eq:phase_shifts}).
The first interaction we considered is a modified Poschl-Teller potential, Eq.~(\ref{eq:mpt}), tuned so that
the scattering length is $a^{nn}=-18.5$ fm and the
effective range is $r_e^{nn}=2.7$ fm.
The other one is based on
the AV18 nucleon-nucleon pairwise interaction \cite{wir95}, which
has been extensively used in QMC simulations of
nucleon systems \cite{car15}.
We chose the neutron-neutron interaction between
particles with antiparallel spins
to be the
$s$-wave
part of AV18.
We fixed the spin-isospin degrees of freedom such that
we have a unpolarized gas of neutrons, hence
the potential interaction becomes spherically symmetrical.
The most important feature of the interaction is that the scattering length $a^{nn}=-18.5$ fm and
effective range $r_e^{nn}=2.7$ fm are correctly described by the potential. In Fig.~\ref{fig:pot} we compare the potential interactions we use for cold gases and neutron matter for $-k_Fa=1$.

\subsection{Wave functions}
\label{sec:wf}

The BCS wave function, which includes pairing explicitly, projected to a fixed number
of particles $N$ (half with spin-up and half with spin-down),
can be written as an antisymmetrized product \cite{bou88}.
Since neither the Hamiltonian or any operators in the quantities we 
calculate flip the spins, we adopt hereafter the convention of primed 
indices 
to denote spin-down particles and
unprimed ones to refer to spin-up particles. Thus, the BCS
wave function 
reduces to
\begin{flalign}
\label{eq:BCS_anti}
&\psi_{\rm BCS}(\textbf{R},S) = \mathcal{A}[ \phi(\textbf{r}_1,s_1, 
\textbf{r}_{1'},s_{1'}) \nonumber \\ 
&\phi(\textbf{r}_2,s_2,
\textbf{r}_{2'},s_{2'})\dots
\phi(\textbf{r}_{N/2},s_{N/2},\textbf{r}_{N/2'},s_{N/2'})],
\end{flalign}
where $\textbf{R}$ is a vector containing the particle positions
$\textbf{r}_i$, $S$ stands for the spins $s_i$,
and the antisymmetrization is over spin-up and spin-down particles 
only \cite{car03}.
This wave function can be 
calculated efficiently as
a determinant.
The $\phi$ are
pairing functions, which have the form
\begin{flalign}
\phi(\textbf{r},s,\textbf{r}',s') = \tilde{\phi}(\textbf{r},
\textbf{r}') \left[\frac{\product{s \ s'}{\uparrow \ \downarrow} - 
\product{s \ s'}{ \downarrow \ \uparrow}}{\sqrt{2}} \right],
\end{flalign}
where we have explicitly included the spin part to impose singlet 
pairing. The assumed expressions for $\tilde{\phi}$ depend on the 
system being studied, see Secs. \ref{sec:wfbulk}, \ref{sec:wfcyl}, and 
\ref{sef:wfvortex}.

The BCS wave function accounts for the long-range behavior.
Short-range correlations are included in the form of
a two-body
Jastrow factor $f(r_{ij'})$, $r_{ij'}=|\textbf{r}_i-\textbf{r}_{j'}|$, 
which accounts for
correlations between antiparallel spins. It is obtained from solutions 
of 
the two-body
Schrödinger-like equation,
\begin{equation}
\left[ -\frac{\hbar^2}{m}\nabla^2 + V(r) 
\right]f(r<d)=\lambda f(r<d),
\end{equation}
where $V(r)$ is specified for cold gases and neutron matter in
Sec.~\ref{sec:scatt}, and
the boundary conditions are $f(r>d)=1$ and $f'(r=d)=0$,
$d$ being a variational parameter, and $\lambda$ is adjusted so that 
$f(r)$ is nodeless. The total trial wave function is written as
\begin{equation}
\psi_{\rm T}(\textbf{R},S)=\prod_{i,j'} f(r_{ij'}) \psi_{\rm BCS}
(\textbf{R},S).
\end{equation}

\subsubsection{Bulk system}
\label{sec:wfbulk}

We employed the same pairing function for the bulk case 
as Ref.~\cite{car03},
\begin{flalign}
\label{eq:bulk}
\tilde{\phi}_{\rm bulk}(\textbf{r},\textbf{r}') = \sum_{n=1}^{n_c} 
\alpha_n e^{i \textbf{k}_{\textbf{n}} \cdot (\textbf{r} - \textbf{r}')} 
+ \tilde{\beta}(|\textbf{r}-\textbf{r}'|),
\end{flalign}
where $\alpha_n$ are variational parameters, and contributions from 
momentum states up to a level $n_c$ are included.
The $\tilde{\beta}$ function describes 
contributions with 
$n>n_c$,
\begin{equation}
\tilde{\beta}(r)=
\begin{cases}
\beta(r)+\beta(L-r)-2\beta(L/2)
 &\text{for } r\leqslant L/2\\
0 &\text{for } r > L/2
\end{cases}
\end{equation}
with
\begin{eqnarray}
\label{eq:beta}
\beta(r) = [1+c b r][1-e^{-dbr}]\frac{e^{-br}}{dbr},
\end{eqnarray}
where $r=|\textbf{r}-\textbf{r}'|$ and $b$, $c$, and $d$ are
variational parameters.
We considered $b$ = 0.5 $k_F$, $d=5$, and $c$ is adjusted so that 
$\partial \tilde{\beta}/\partial r=0$ at $r=0$.
This functional form of $\beta(r)$ describes the short-distance
(high-momentum)
correlation of particles with antiparallel spins.

\subsubsection{Cylinder}
\label{sec:wfcyl}

The free-particle solution of the Schr\"odinger equation
in a cylinder or radius ${\cal R}$, height ${\cal L}$,
finite at $\rho=0$, and with periodic conditions
along the $z$-axis is
\begin{equation}
\label{eq:spcyl}
\Phi_{n\nu p}(\rho,\varphi,z)={\cal N}_{\nu p}
J_\nu(k_{\nu p}\rho)\exp\left[i(k_z z+\nu\varphi)\right],
\end{equation}
where ${\cal N}_{\nu p}$ is a normalization constant,
$J_\nu$ are Bessel functions, $k_{\nu p}=j_{\nu p}/{\cal R}$,
$j_{\nu p}$ is the $p$-th zero of $J_\nu$,
and $k_z=2\pi n/{\cal L}$. The eigenvalues are
$E_{n\nu p}=\hbar^2(k_{\nu p}^2+k_z^2)/(2m)$.
The quantum numbers $n$ and $\nu$ can take the values
0,$\pm 1,\pm 2,\cdots$, and $p=1,2,\cdots$.

The pairing function for the cylinder geometry is constructed using the 
single-particle orbitals of Eq.~(\ref{eq:spcyl}) coupled with
their time-reversed counterparts. 
This ansatz has been used before in the unitary Fermi gas \cite{mad16}.
We assume the pairing 
function to be 
\begin{flalign}
\label{eq:cyl}
\tilde{\phi}_{\rm cyl}(\textbf{r},\textbf{r}') = \sum_{q=1}^{q_c} 
\tilde{\alpha}_q
\mathcal{N}_{\nu p}^2 J_\nu\left(\frac{j_{\nu p}}
{\mathcal{R}}\rho\right) J_\nu\left(\frac{j_{\nu p}}{\mathcal{R}}\rho' 
\right)
\nonumber\\
\times
e^{i \nu (\varphi-\varphi')}e^{i k_z (z-z')}+ 
\bar{\beta}(\textbf{r},
\textbf{r}'),
\end{flalign}
where
the $\tilde{\alpha}_q$ are variational parameters, and
$q$ is a label for the cylinder momentum shells, such that different states with the 
same energy have the same variational parameter.
The 
$\bar{\beta}$ function is a modification of $\tilde{\beta}$
such that the hard wall boundary 
condition is met,
\begin{equation}
\label{eq:betacyl}
\bar{\beta}(\textbf{r},\textbf{r}')=
\begin{cases}
{\cal N}_{01}^2 J_0\left(\frac{j_{01}\rho}{{\cal R}}\right) 
J_0\left(\frac{j_{01}\rho'}{{\cal R}}\right)\times \\
\left[ 
\beta(r)+\beta(2{\cal R}-r)-2\beta({\cal R})\right]
 &\text{for } r\leqslant {\cal R}\\
0 &\text{for } r > {\cal R}
\end{cases}
\end{equation}
and $\beta$ is given by 
Eq.~(\ref{eq:beta}).

\subsubsection{Vortex}
\label{sef:wfvortex}

The vortex line excitation is accomplished by considering pairing orbitals 
which are eigenstates of $L_z$ with eigenvalues $\pm \hbar$. This is 
achieved by coupling single-particle states with angular quantum 
numbers $\nu$ differing by one.
Explicitly, we are considering $(n,\nu,p)$ paired with $(-n,-\nu+1,p)$,
such that the pairing orbitals take the 
form
\begin{flalign}
\label{eq:vortex}
&\tilde{\phi}_{\rm vortex}(\textbf{r},\textbf{r}') = \sum_{q=1}^{q_c} 
\bar{\alpha}_q
\mathcal{N}_{\nu p} \mathcal{N}_{\nu-1 ; p}\times \nonumber \\
&\left\{ J_\nu\left(\frac{j_{\nu p}}{\mathcal{R}}\rho\right) 
J_{\nu-1}\left(\frac{j_{\nu-1;p}}{\mathcal{R}}\rho' \right) e^{i (\nu 
\varphi- (\nu-1) \varphi')}
e^{ik_z(z-z')}
\right. \nonumber \\
&+ \left. J_\nu\left(\frac{j_{\nu p}}{\mathcal{R}}\rho'\right) 
J_{\nu-1}\left(\frac{j_{\nu-1;p}}{\mathcal{R}}\rho \right) e^{i (\nu 
\varphi'- (\nu-1) \varphi)} e^{ik_z(z'-z)}\right\},
\end{flalign}
where $q$ is a label for the vortex shells, and $\bar{\alpha}_q$ are 
variational parameters. 
Equation~(\ref{eq:vortex}) is symmetric under interchange of the prime and 
unprimed coordinates, as required for singlet 
pairing.

\subsection{Quantum Monte Carlo}
\label{sec:qmc}

The Hamiltonian of the two-component Fermi gas,
or spin-up/spin-down neutron matter,
is given by
\begin{equation}
H=-\frac{\hbar^2}{2m}\left[ \sum_{i=1}^{N_\uparrow} \nabla_i^2 + 
\sum_{i=j'}^{N_\downarrow} \nabla_{j'}^2 \right] + \sum_{i,j'} 
V(r_{ij'}),
\end{equation}
with $N=N_\uparrow+N_\downarrow$.
The diffusion Monte Carlo (DMC) method projects out the lowest-energy state of $H$
present in a
initial 
state $\psi_T$ (obtained from variational Monte Carlo 
simulations). The propagation, in imaginary time 
$\tau$, can be written as
\begin{equation}
\psi(\tau)= e^{-(H-E_T)\tau}\psi_T,
\end{equation}
where $E_T$ is an energy offset. In the $\tau\to\infty$ limit, only the 
lowest energy component $\Phi_0$ survives
\begin{equation}
\lim_{\tau \to \infty} \psi(\tau)=\Phi_0.
\end{equation}
The imaginary time evolution can be written in the integral form
\begin{equation}
\label{eq:green}
\psi(\textbf{R},\tau)=\int d\textbf{R}' G(\textbf{R},\textbf{R}',
\tau)\psi_T(\textbf{R}'),
\end{equation}
where $G(\textbf{R},\textbf{R}',\tau)$ is the Green's function 
associated with $H$. We solve an importance sampled version of 
Eq.~(\ref{eq:green}) iteratively, using the Trotter-Suzuki 
approximation to evaluate $G(\textbf{R},\textbf{R}',\tau)$, which 
requires the time steps $\delta \tau=\tau/N$ to be small. We circumvent the 
fermion-sign problem by using the fixed-node approximation, which 
restricts transitions across a nodal surface defined
by $\psi_T$, making 
our estimates of energy expectation values upper bounds.
For a detailed explanation of the algorithm, the reader is
referred to Ref.~\cite{fou01} and references therein.

The direct calculation of the expectation value of an operator $O(\textbf{R})$ from $\Phi_0(\textbf{R})$ corresponds to the mixed estimator
\begin{eqnarray}
\expected{O(\textbf{R})}_m = \frac{\bra{\Psi_T(\textbf{R})}O(\textbf{R})\ket{\Phi_0(\textbf{R})}}{\product{\Psi_T(\textbf{R})}{\Phi_0(\textbf{R})}},
\end{eqnarray}
which is exact only when $O$ commutes with the Hamiltonian $H$.
There are several methods to compute expectation values of quantities, such as the density, that do not commute with $H$. One of them is the extrapolation method where the results of diffusion and variational simulations are combined.
However, the accuracy of the extrapolation method relies completely on the trial wave function. Moreover, even in the case of accurate trial wave functions, the bias of the extrapolated estimator is difficult to
calculate.
For these reasons we used the forward walking method, which is discussed in detail in Ref.~\cite{cas95}, to evaluate the
density profiles. This method relies on the calculation of the asymptotic offspring of walkers coming from the branching term to compute
the exact estimator,
\begin{eqnarray}
\label{eq:exact_est}
\expected{O(\textbf{R})}_e = \frac{\bra{\Phi_0(\textbf{R})}O(\textbf{R})\ket{\Phi_0(\textbf{R})}}{\product{\Phi_0(\textbf{R})}{\Phi_0(\textbf{R})}}.
\end{eqnarray}

The variational parameters
in Eqs.~(\ref{eq:bulk}), (\ref{eq:cyl}), and (\ref{eq:vortex}) were 
determined using the
linear method \cite{tou07}. In this method, parameter
variations are found by diagonalizing a non-symmetric
estimator of the Hamiltonian matrix in the basis of the wave function and its derivatives with respect to the parameters.
We also adopted the heuristic procedure of Ref.~\cite{Contessi2017}, which suppresses instabilities that arise from the non-linear dependence of the wave function on the variational parameters.

\section{Results}
\label{sec:res}

Comparison between cold atom systems and low-density neutron matter
is achieved by expressing energies (per particle) in units of the
free Fermi gas energy, see Eq.~(\ref{eq:EFG}), and distances in units
of $k_F^{-1}$.
For the cold gases systems we keep the effective range fixed at
$k_F r_e=0.05$, which is much smaller than the interparticle spacing and
the scattering lengths involved in the simulations.
The number density is kept constant at $n=k_F^3/(3\pi^2)$. For bulk systems
this corresponds to $n=N/L^3$, and for cylindrical containers
$n=N/(\pi{\cal R}^2{\cal L})$. The interaction strengths we considered
for cold gas systems are
$-k_Fa=\{0.5,1.0,2.0,3.3,5.0\}$, while we do not include
the $-k_Fa=0.5$ case for neutron matter because it is extremely dilute,
and Friedel oscillations prevent any meaningful analysis of the density
profiles.

\subsection{Energy}
\label{sec:energy}

\subsubsection{Ground state energy}
\label{sec:gscyl}

\begin{figure}[!htb]
\centering
\includegraphics[angle=-90,width=\linewidth]{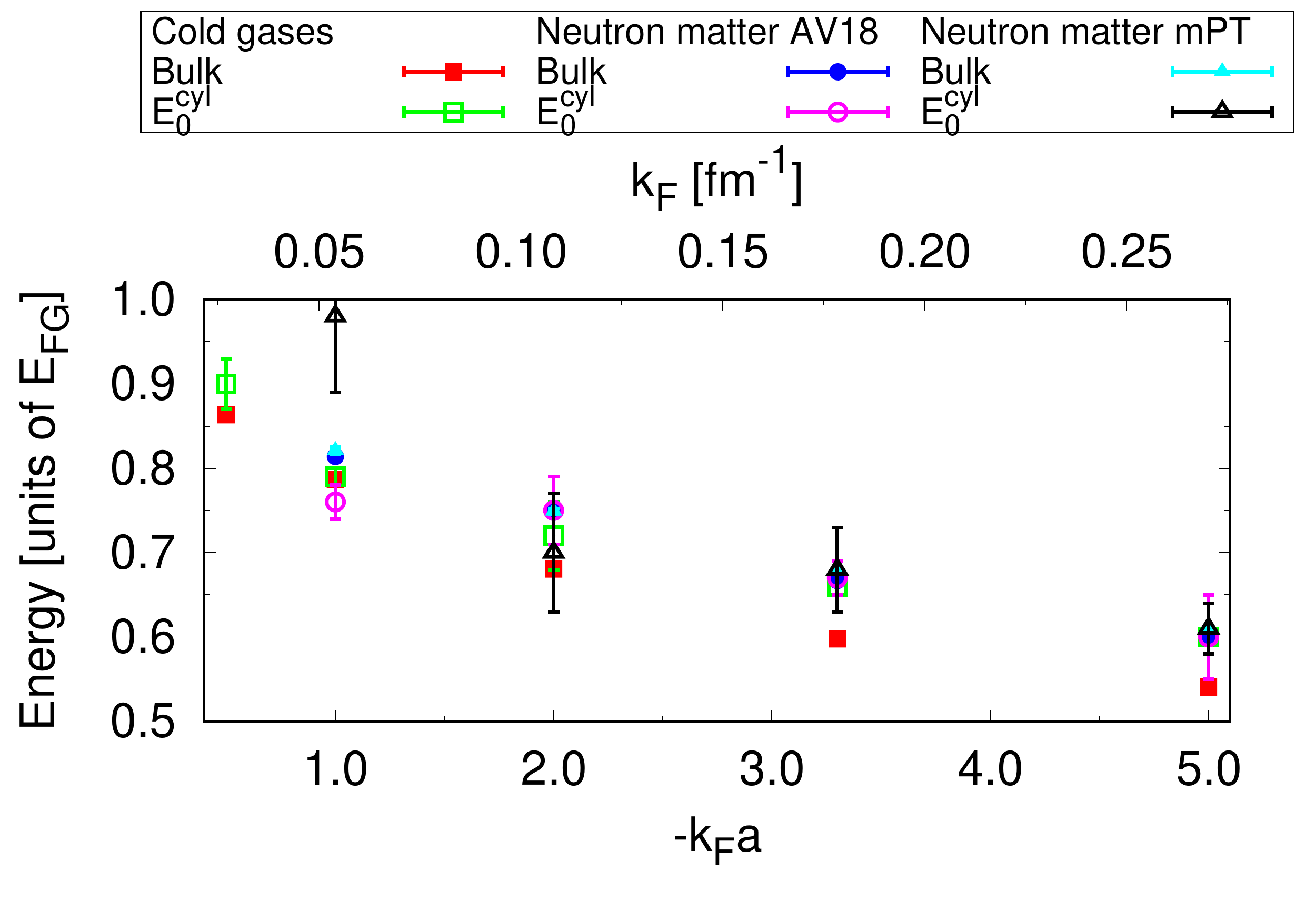}
\caption{(Color online) Equation of state for cold atoms and
low-density neutron matter.
The bulk energies per particle for the cold gases, closed (red) squares, were obtained
using the modified Poschl-Teller potential with $k_F r_{\rm e}=0.05$.
For neutron matter,
the bulk energies per particle using the $s$-wave part of AV18 are denoted by
closed (blue) circles, and the model using a modified
Poschl-Teller potential with $a^{nn}=-18.5$ fm and $r^{nn}_{\rm e}=2.7$ fm
is represented by
closed (cyan) triangles.
We also plot the fitted parameters $E_0^{\rm cyl}$ of Eq.~(\ref{eq:ERL}) for cold gases and neutron matter ($s$-wave part of AV18 and modified
Poschl-Teller) with open symbols:
(green) squares,
(magenta) circles,
(black) triangles,
respectively.
In the top $x$ axis we plot the corresponding $k_F$ for neutron matter.
}
\label{fig:eos}
\end{figure}

The ground state energy per particle of the bulk systems for several values of
$k_F a$ was calculated using the pairing function of 
Eq.~(\ref{eq:bulk}), and the
results are shown in Table~\ref{tab:energy} and also in 
Fig.~\ref{fig:eos}.
The energy per particle of the cold atoms systems is lower
than the neutron matter systems, for the same value of $k_Fa$, in
accordance with previous simulations. In fact, our results for
cold atoms are lower than those reported in Ref.~\cite{Gezerlis:2008}
because we chose a smaller effective range, $k_F r_e=0.05$, than
the value employed by them. As for the bulk energies comparing
the two models for the neutron matter interactions,
the values obtained using the modified
Poschl-Teller potential are slightly larger than the ones using the
$s$-wave part of AV18, although the relative
difference is 2\% at most.

\begin{table*}[!htbp]
\caption{Bulk energies per particle and the parameters $E_0^{\rm cyl}$ and $\lambda_S$ fitted to the functional form of Eq.~(\ref{eq:ERL}). The bulk energies and $E_0^{\rm cyl}$ are reported
in units of the free Fermi gas energy, $E_{FG}$ (see Eq.~(\ref{eq:EFG})), while $\lambda_S$ is reported in units of $E_{FG} k_F^{-2}$.}
\begin{tabular}{|c|c|c|c||c|c|c|c|c|c|}
\hline
&\multicolumn{3}{c||}{\textbf{Cold gases}} & \multicolumn{6}{c|}{\textbf{Neutron matter}}\\ \hline
&\multicolumn{3}{c||}{} & \multicolumn{3}{c|}{$s$-wave AV18}
& \multicolumn{3}{c|}{modified Poschl-Teller} \\ \hline
\multicolumn{1}{|c|}{$-k_Fa$} & bulk & $E_0^{\rm cyl}$ & $\lambda_S$ & bulk & $E_0^{\rm cyl}$ & $\lambda_S$ & bulk & $E_0^{\rm cyl}$ & $\lambda_S$ \\ \hline
0.5 & 0.8636(1) & 0.90(3) & 95(15) &  &  &  &  &  &  \\ \hline
1.0 & 0.7864(2) & 0.79(1) & 99(7) & 0.814(5) & 0.76(2) & -78(9) & 0.821(4) & 0.98(9) & 204(54) \\ \hline
2.0 & 0.6806(2) & 0.72(4) & 70(19) & 0.748(2) & 0.75(4) & -242(10) & 0.749(3) & 0.70(7) & 117(30) \\ \hline
3.3 & 0.5979(2) & 0.66(1) & 58(4) & 0.667(2) & 0.67(2) & -365(70) & 0.681(2) & 0.68(5) & 85(24) \\ \hline
5.0 & 0.5407(2) & 0.60(1) & 64(5) & 0.598(2) & 0.60(5) & -445(80) & 0.608(1) & 0.61(3) & 85(14) \\ \hline
\end{tabular}
\label{tab:energy}
\end{table*}

We used the pairing functions of Eq.~(\ref{eq:cyl}) to calculate 
the ground state energy of the cylindrical systems for
$N=\{78,80,82,84,86\}$. Then we fitted the results to the functional
form of Eq.~(\ref{eq:ERL}), and we report the parameters $E_0^{\rm cyl}$ and $\lambda_S$ in Table~\ref{tab:energy}.
Ideally we would like to have $E_0^{\rm cyl}$ match the bulk
value for every interaction strength, meaning that we can
separate the ground state energy of the fermionic systems into a bulk
component and a surface term. For both cold gases and low-density 
neutron matter, and most interaction strengths, the results
are within the error bars.
For the $s$-wave part of the AV18 model the agreement is quite good. The relative
difference does not exceed 11\%, and most values agree within error bars.
It is worth pointing out that the values of $\lambda_S$ are negative
for these systems due to the repulsive core of the interaction, see
Fig.~\ref{fig:pot}, a feature that is absent in the purely attractive
potentials employed in the other cases.
For the modified Poschl-Teller potential, the fitting procedure yielded
larger errors. Also, the results for $-k_Fa=1$ do not follow the trend, most
probably due to the diluteness of the system.
In Fig.~\ref{fig:eos} we compare the
bulk energies of cold gases and neutron matter with the corresponding
values of $E_0^{\rm cyl}$.

In Sec.~\ref{sec:scatt} we presented the potential interaction
used for neutrons of antiparallel spins, and we set the interaction
between particles of the same spin to zero. In doing so,
we neglected the interaction of the $M=\pm 1$ triplet states.
Previous QMC simulations of bulk low-density neutron matter \cite{Gezerlis:2008}, using
a similar formalism to ours,
found that, perturbatively, corrections for the
artificial attraction in the $M=0$ triplet state account
for 10\% of the total energy in the $-k_F a=10$ case. The corrections
become even lower for lower densities, such that in the range
considered in this work they are of order of a few percent.
Later calculations \cite{gez10} compared results using the pure $s$-wave interaction
with the AV4'\cite{Wiringa:2002} potential, which yielded $\sim$ 7\%
difference for $-k_F a=10$,  $\sim$ 1\% for $-k_F a=5$, and
essentially the same results for lower densities.
These results in the bulk neutron matter systems
justify our approach because, besides vanishing small
corrections to the total energy as the density is lowered,
one of
our goals is to calculate the vortex excitation energy, which
is an energy difference, and thus the corrections is expected to cancel.

\subsubsection{Vortex excitation energy}

\begin{figure}[!htb]
\centering
\includegraphics[angle=-90,width=\linewidth]{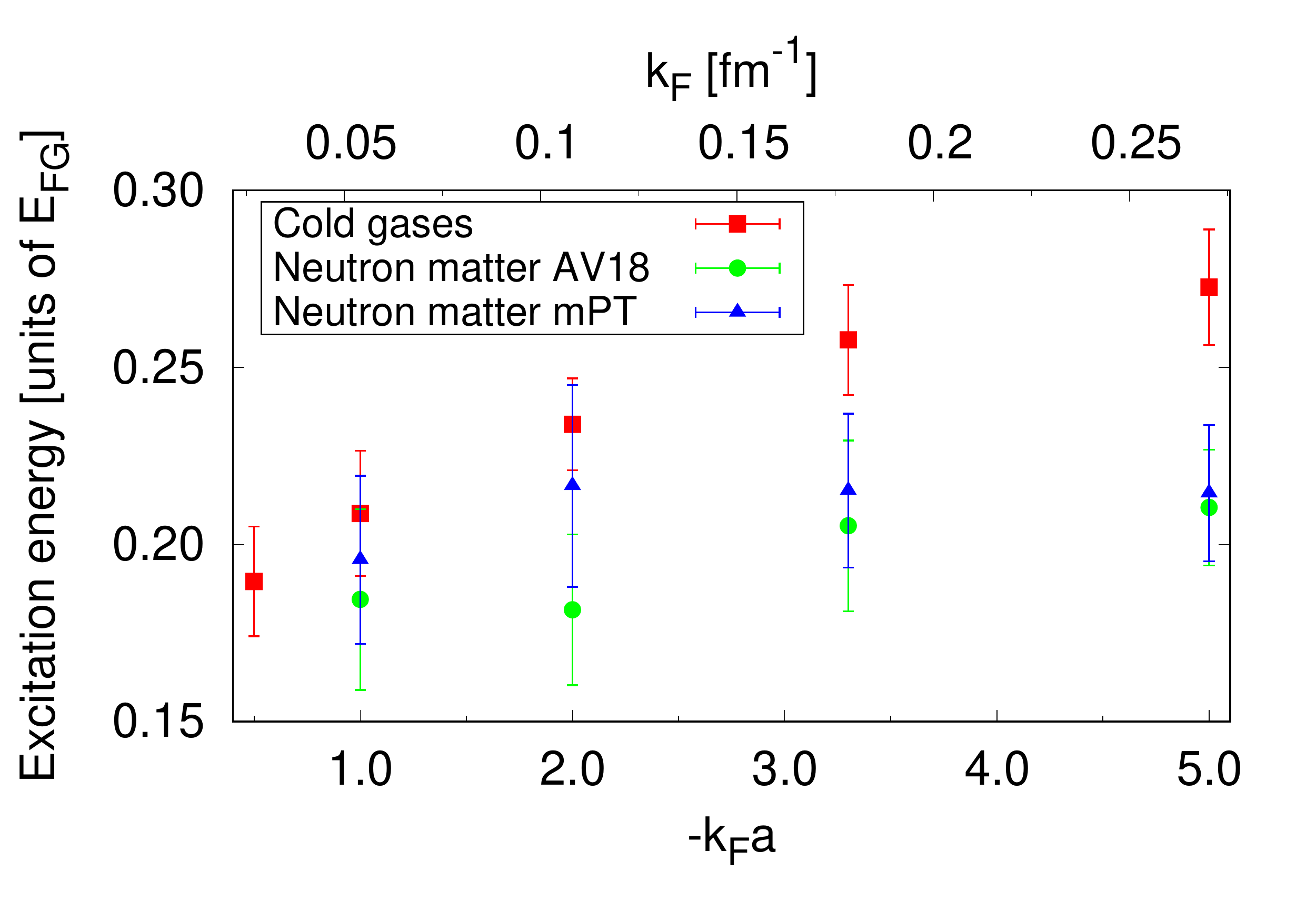}
\caption{(Color online) Excitation energy per particle as a function
of the interaction strength for both cold gases and neutron matter.
The (red) squares denote the results for cold gases, i.e.,
using the modified Poschl-Teller potential with $k_F r_{\rm e}=0.05$.
For neutron matter,
the results using the $s$-wave part of AV18 are denoted by (green) circles, and the model using a modified
Poschl-Teller potential with $a^{nn}=-18.5$ fm and $r^{nn}_{\rm e}=2.7$ fm
is represented by (blue) triangles. 
In the top $x$ axis we plot the corresponding $k_F$ for neutron matter.
We can see that the excitation energies are comparable for $-k_F a=1$, however,
when the density (or $-k_F a$) increases, they start to differ.
}
\label{fig:exc}
\end{figure}

The energy of the systems with a vortex line was calculated using
the pairing functions of Eq.~(\ref{eq:vortex}).
The excitation energy was computed using the energy difference between
those systems and the ground state of the cylinder. The results were
averaged for $N=\{78,80,82,84,86\}$. Figure~\ref{fig:exc} shows the excitation
energy for low-density neutron matter and cold atoms as a function
of $k_F a$.

Although several results are within error bars, we can see that
for $-k_Fa=1$ the vortex excitation energy for cold gases and neutron matter
(both models) is comparable. As the interaction strength increases,
we can clearly see that
the
excitation energy is higher for cold gases systems compared to
low-density neutron matter.
The results for neutron matter, using both models, seem to be much less dependent on the interaction strength for this density regime.
As was the case in the previous section, the errors associated with
the modified Poschl-Teller potential for neutron matter are larger than the
other two cases, however, it is still possible to see that the results
for the two neutron matter models are close.

\subsection{Density profiles}
\label{sec:dens}

\begin{figure}[!htb]
  \includegraphics[angle=-90,width=\linewidth]{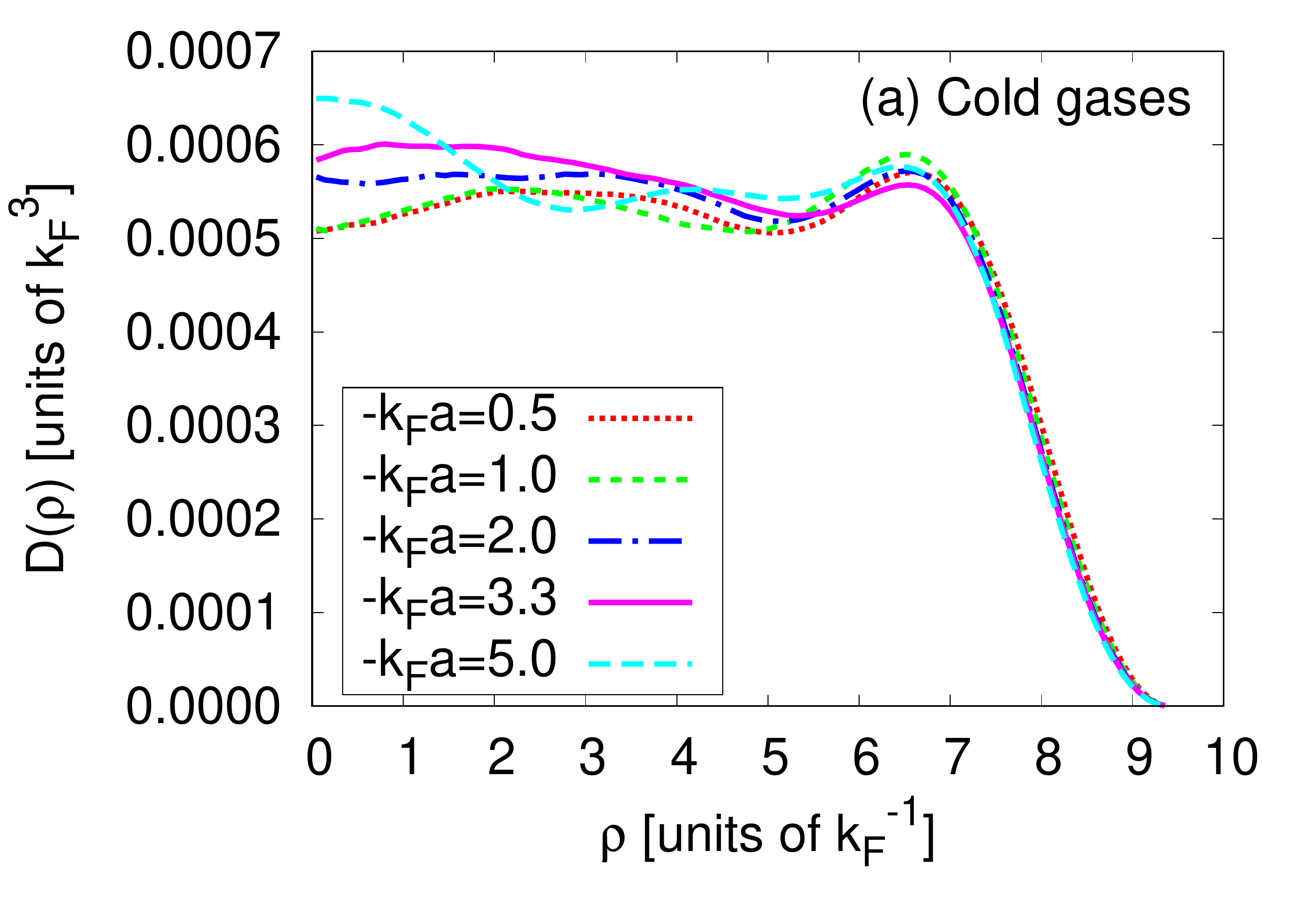}
  \includegraphics[angle=-90,width=\linewidth]{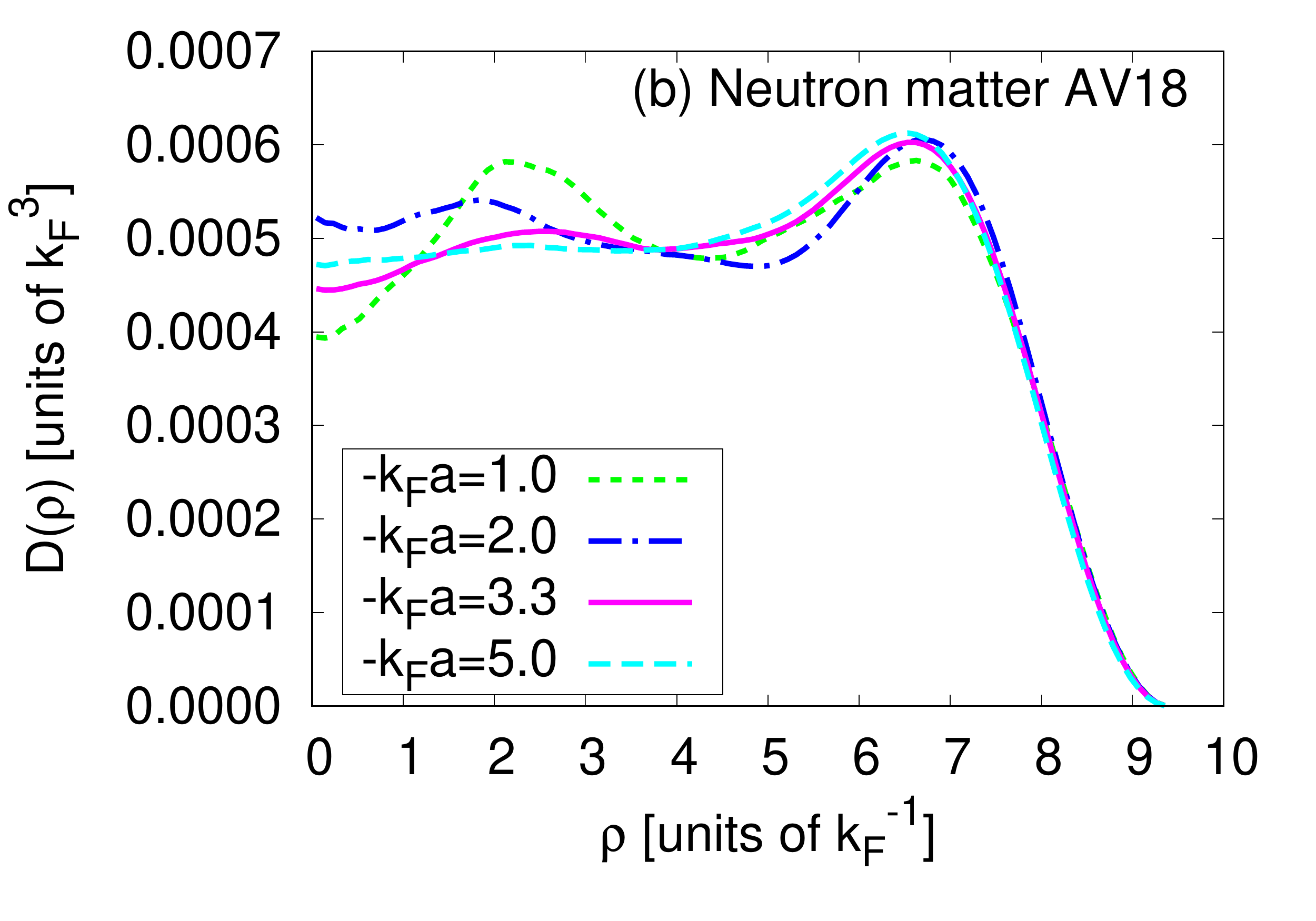}
  \includegraphics[angle=-90,width=\linewidth]{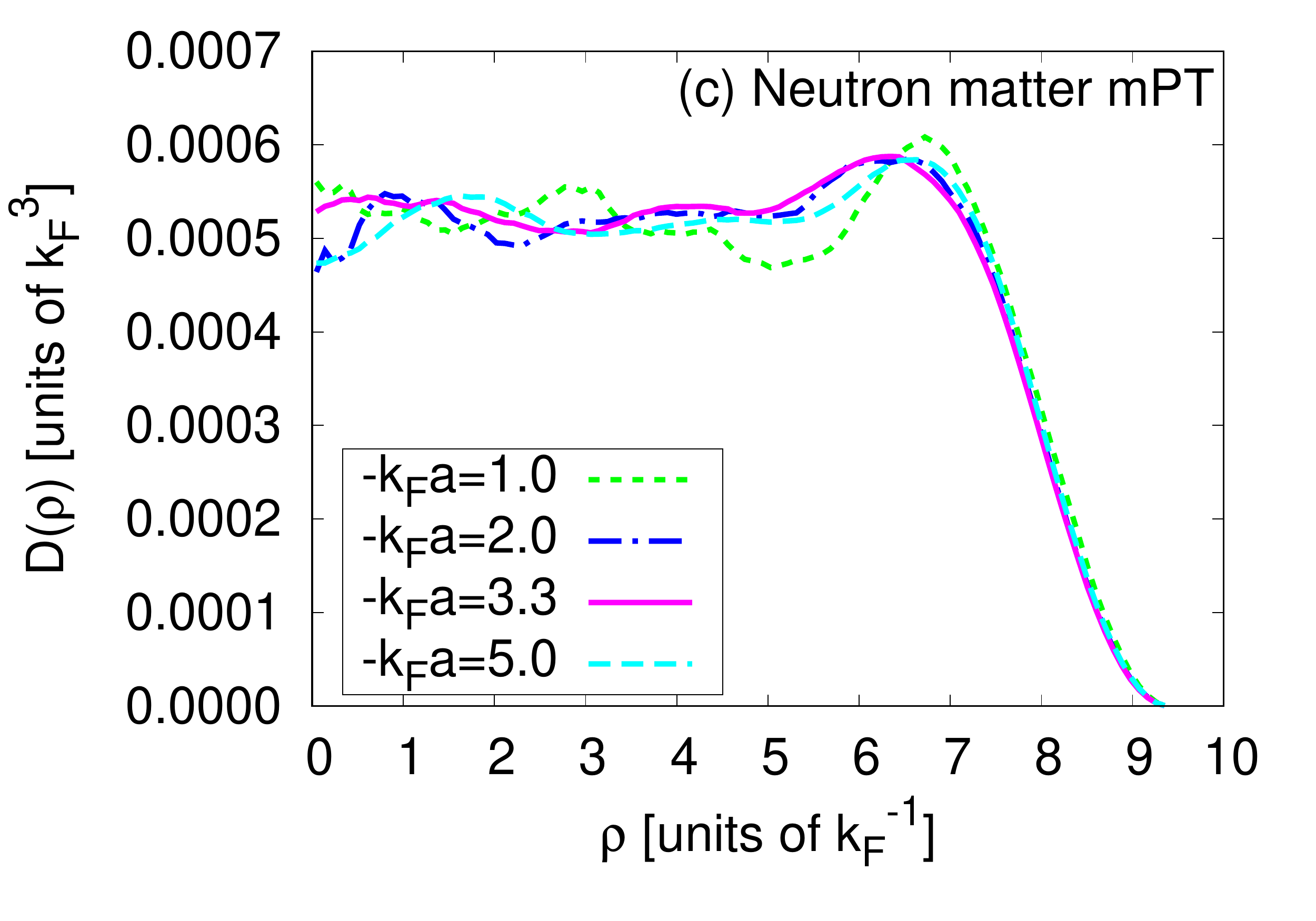}
\caption{(Color online)
Density profile of the ground state as a function
of the radial coordinate $\rho$ for $N=84$
for (a) cold gases, (b) neutron matter using the $s$-wave part of AV18,
and (c) the modified Poschl-Teller potential.
The interaction strengths $-k_Fa=\{0.5,1.0,2.0,3.3,5.0\}$
correspond to the
short-dashed (red) line,
dashed (green) line,
dashed-dotted (blue) line,
solid (magenta) line, and
long-dashed (cyan) line, respectively.
}
\label{fig:gsdens}
\end{figure}

The density profile ${\cal D}(\rho)$ was calculated averaging the
angular ($\varphi$) and axial ($z$) directions.
We chose a normalization such that
\begin{eqnarray}
\int_V d^3r {\cal D}(\rho)=1,
\end{eqnarray}
where the integral is over the volume $V=\pi{\cal R}^2{\cal L}$ of the cylinder. We show our results for the ground state density of the cylindrical container in Fig.~\ref{fig:gsdens}.
The results for cold gases, Fig.~\ref{fig:gsdens}(a), follow a similar trend, with
the exception of $-k_Fa=5.0$, the largest interaction strength considered. Nonetheless, the Friedel oscillations are much smoother
than in the neutron matter systems, Figs.~\ref{fig:gsdens}(b) and \ref{fig:gsdens}(c).
The results using the $s$-wave of the AV18 model show
a very pronounced oscillation near
$\approx$ 2.0$k_F^{-1}$ for $-k_Fa=1.0$, and it
is less intense for stronger interactions.

\begin{figure}[!htb]
\includegraphics[angle=-90,width=\linewidth]{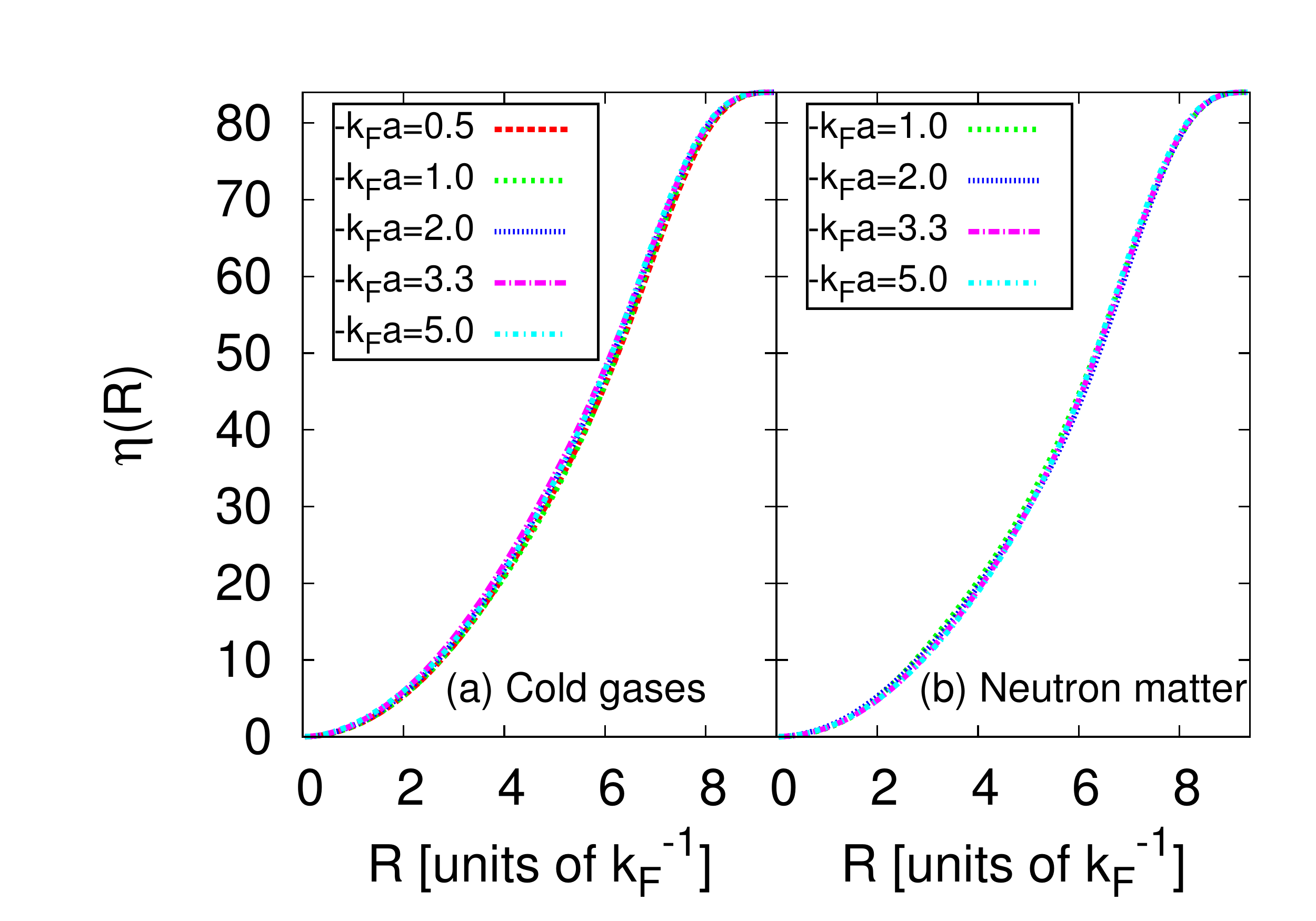}
\caption{(Color online)
Particle number $\eta$ a distance $R$ from the $z$ axis, see Eq.~(\ref{eq:intdens}),
for the ground state and $N=84$
for (a) cold gases and (b) neutron matter using the $s$-wave part of AV18.
The legend conventions are the same as the ones employed in
Fig.~\ref{fig:gsdens}. The deviations between the behavior
of different interaction strengths, or cold gases and neutron
matter, are very small.
Also, the differences between the two models for the neutron-neutron
interactions are so minute that we chose to plot only one of them.
An inspection of Fig.~\ref{fig:gsdens} reveals a characteristic
behavior of the density, due to the presence of hard walls,
at $\rho\approx 6.0 k_F^{-1}$.
For $R\simeq 6.0 k_F^{-1}$, $\eta \gtrsim$ 45, meaning that
we have approximately this number of particles in the ``bulk''
portion of the cylinder, where effects of the hard walls are
mitigated.
}
\label{fig:dens_int}
\end{figure}

The hard wall condition introduces a characteristic density
behavior close to it as it was discussed in Sec.~\ref{sec:cyl},
and as seen in Fig.~\ref{fig:gsdens}. We were able to separate
two contributions to the ground state energy of the cylindrical
systems, which we identified as bulk and surface terms, Sec.~\ref{sec:gscyl}. However, this analysis requires that
there is a sufficient number of particles in the central region
of the cylinder, away from the walls.
To this end, we define the particle number $\eta(R)$ a distance $R$ from the $z$ axis,
\begin{eqnarray}
\label{eq:intdens}
\eta(R)=N\int_{0}^{{\cal L}}dz\int_{0}^{2\pi}d\varphi
\int_0^R d\rho \rho {\cal D}(\rho),
\end{eqnarray}
such that $\eta(R={\cal R})=N$.
In Fig.~\ref{fig:dens_int} we plot $\eta(R)$ for cold gases and neutron
matter systems using $N=84$ particles, which show essentially the same behavior, independently
of the interaction strength.
Figure~\ref{fig:gsdens}
suggests that the hard walls affect the systems at
$\rho\gtrsim 6.0 k_F^{-1}$. 
As we can see in Fig.~\ref{fig:dens_int},
$\eta(\sim 6.0 k_F^{-1}) \gtrsim$ 45, meaning that
we have approximately this number of particles in the ``bulk''
portion of the cylinder. For systems with a vortex line this number is lower, $\approx$ 42.
Previous QMC simulations of bulk properties have employed
$N=$38 \cite{car03}
and $N=$40 \cite{for11}, hence the number of particles we have in
the center of the cylinder is larger than in those bulk calculations.

\begin{figure*}[!htb]
\includegraphics[width=\linewidth]{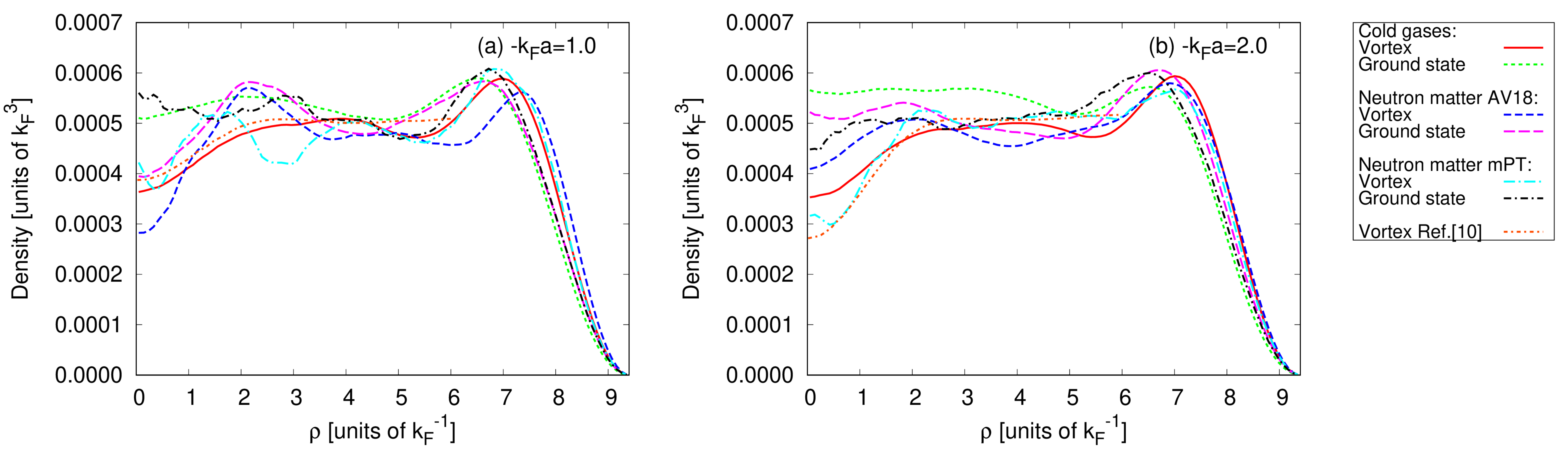}
\caption{(Color online)
Density profile of the vortex and ground state of cold gases
and neutron matter as a function
of the radial coordinate $\rho$ for $N=84$.
Panel (a) corresponds to an
interaction strength of $-k_Fa=1.0$, and (b) to $-k_Fa=2.0$.
The cold gases vortex and ground state profiles are represented by
continuous (red) lines and
short-dashed (green) lines, respectively. 
The results for neutron matter using the $s$-wave part of AV18
are plotted with
dashed (blue) lines,
and long dashed (magenta) lines, while the modified Poschl-Teller model
is represented by
long dashed-dotted (cyan) lines and short dashed-dotted black lines.
We also plot the results from Ref.~\cite{sen06} for cold gases, using dotted short dashed-dotted (orange) lines, where
we changed their normalization so that both systems have the same number of
particles inside the $\rho\leqslant 6.0k_F^{-1}$ region.
}
\label{fig:vortexm1m2}
\end{figure*}

\begin{figure*}[!htb]
\includegraphics[width=\linewidth]{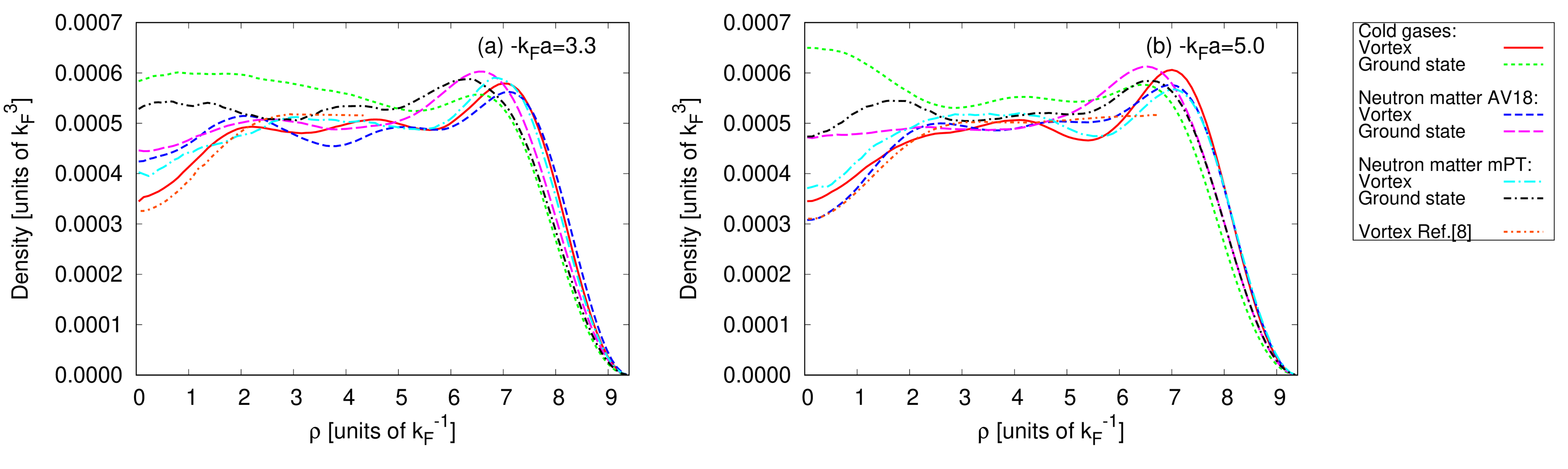}
\caption{(Color online)
Density profile of the vortex and ground state of cold gases
and neutron matter as a function
of the radial coordinate $\rho$ for $N=84$.
Panel (a) corresponds to an
interaction strength of $-k_Fa=3.3$, and (b) to $-k_Fa=5.0$.
The legend conventions for our results are the same as the ones employed
in Fig.~\ref{fig:vortexm1m2}.
We compare our results with Ref.~\cite{yu03} for neutron matter, short dashed-dotted (orange) lines, where
we changed their normalization so that both systems have the same number of
particles inside the $\rho\leqslant 24.5$ fm region.
}
\label{fig:vortexm3m5}
\end{figure*}

In Figs.~\ref{fig:vortexm1m2} and \ref{fig:vortexm3m5} we plot
the density profiles for $-k_Fa=\{0.5,1.0,2.0,3.3,5.0\}$ of the
ground and vortex line states for cold atoms and neutron matter.
We compared the density profiles of cold gases for $-k_Fa=1.0$ and 2.0,
Fig.~\ref{fig:vortexm1m2},
with the
Bogoliubov-de Gennes
calculations of Ref.~\cite{sen06}.
They used a different geometry than ours, so to compare the
results we changed their normalization to match our number of particles
in the $\rho\leqslant 6.0k_F^{-1}$ region of the cylinder.
Their results are closer to ours in the $-k_Fa=1.0$ case, as expected.
In the low-density neutron matter case, we compared
our results for $-k_Fa=3.3,5.0$ with the mean-field
results of Ref.~\cite{yu03}, see Fig.~\ref{fig:vortexm3m5}.
In a similar fashion to what we did in the cold atoms case, we matched
the normalizations to ensure the same number of particles in the
$\rho\leqslant 24.5$ fm region.

A direct comparison of the density profiles for cold gases and neutron matter,
or the two models we used for neutron matter,
is difficult due to the different position of the oscillations in the profiles. 
However,
a quantity of interest in both rotating
superfluid cold gases systems
and neutron matter is the density depletion at the vortex core, which depends
only on the density near the origin.
In the BCS limit the density should be close to the
ground state one, while in the BEC regime
the core should be completely depleted.
We found that,
in the cold gases case, the ratio of the density at
$\rho=0$ for the system with a vortex line and the
ground state of the cylindrical container decreases from
72\% ($-k_Fa=0.5$) to 53\% ($-k_Fa=5.0$), with
the values 71\%, 62\%, and 59\% for $-k_Fa=1.0,2.0,3.3$, respectively.
The mean-field calculation of Ref.~\cite{sen06} finds a much higher
density close to the BCS limit, 94\% at $-k_Fa=1.0$. However,
their result for $-k_Fa=2.0$ is comparable with ours, 60\%.
For low-density neutron matter using the modified Poschl-Teller potential, we found a density at the core
of approximately
75\%, 71\%, 76\%, and 78\% of the ground state density for $-k_Fa=1.0$, 2.0,3.3, and 5.0, respectively.
The $s$-wave part of AV18 model for neutron matter yields
75\% of the ground state density for $-k_Fa=1.0,2.0$.
For $-k_Fa=5.0$ this ratio is 65\%, close to the value of 60\% of
Ref.~\cite{yu03}. For $-k_Fa=3.3$ we see a small depletion, but that is
an artifact of the density oscillations near the origin, so we chose
not to include this interaction strength in the density at the core
discussion.

\section{Summary and outlook}
\label{sec:sum}

In this work we compare properties of vortices in
low-density neutron matter
and cold atomic dilute Fermi
gases. Our goal is not to show that
they are identical, but rather to draw a parallel between their properties
such that measuring quantities in cold Fermi gases can help to
constrain properties of vortices in neutron matter, as previously 
done for the ground state~\cite{Gandolfi:2015}.

Although the ground-state energies per particle of the bulk systems were 
lower
for cold gases than for neutron matter, for a given $k_Fa$,
the difference becomes smaller as we move toward more dilute 
systems. This was the main motivation to expect that vortices in the
low-density regime show a duality between cold gases and neutron matter.
The excitation energy for the formation of a vortex line is comparable
when the density is low enough.
However, it is higher for cold
gases than in neutron matter, so that must be accounted for
when comparing the
two systems.

We chose to analyze the density depletion at the vortex core, because it only
depends on the density behavior close to the axis of the cylinder, away from the
hard walls.
Again, we found an agreement between the values for very low densities, although
the density at the vortex core tends to remain close to 75\% of the
ground-state density for neutron matter,
whereas we can clearly see it dropping from 72\% to
53\% for cold gases.

We found an excellent agreement when comparing the two models we employed for the neutron-neutron interactions.
It seems remarkable that two potentials of completely different shapes, see Fig.(\ref{fig:pot}), give us the same physical properties.
However, the fact that they have the same scattering length and effective
range is the key feature. This indicates that
the low-energy limit of Eq.~(\ref{eq:phase_shifts}) is also valid
for low-density neutron matter.

Our results can help to relate cold atom experiments
with properties of low-density neutron matter.
The extraction of bulk properties from experiments is extremely 
difficult
when they employ harmonic traps. However, 
box-like traps \cite{gau13} have been successfully implemented
in Bose systems,
and they can help pave the way to determining the equation of 
state for cold gases. That, in turn, could be contrasted with
Fig.~\ref{fig:eos} to constraint the low-density neutron matter
equation of state.
Vortices in fermionic gases on both BCS and BEC sides of the crossover, 
and also at unitarity, have been observed \cite{zwi05}.
%Hopefully our results for the excitation energy, Fig.~\ref{fig:exc},
%can also be used to draw a parallel between those two systems.
%Our density profile results also suggest that previous mean-field 
%calculations
%of both cold gases \cite{sen06} and neutron matter \cite{yu03}
%tend to overestimate the density depletion at the vortex core. 

Our approach is valid for the low-density regime of neutron matter.
However, it would be interesting to investigate vortex properties at higher densities.
In Sec.~\ref{sec:gscyl} we discussed possible corrections
to account for our choice of neutron-neutron interaction 
potential.
We showed that they would be small in the bulk case, thus
justifying our approach, but they increase with the density.
Instead of carefully including corrections, it seems
more promising to consider realistic nuclear Hamiltonians.
There are calculations using auxiliary-field diffusion
Monte Carlo (AFDMC) \cite{gan08,gan09} where bulk properties of neutron matter are calculated, at higher densities than in this present work, using realistic nuclear Hamiltonians. A possible extension of our work is to generalize the wave functions we presented by including spin correlations, and perform AFDMC simulations.
The comparison of the results using both methods should enlighten how important spin correlations are when describing low-density neutron matter.  

\begin{acknowledgments}
We thank Alessandro Lovato for the useful discussions.
This work was supported by
the São Paulo Research Foundation (FAPESP)
under the grant 2018/09191-7, and by 
the National Science
Foundation under the grant PHY-1404405.
The work of S.G. was supported by the NUCLEI
SciDAC program, by the U.S. DOE under contract
DE-AC52-06NA25396,
by the LANL LDRD program, and by
the DOE Early Career Research Program.
This work used the Extreme Science and Engineering Discovery Environment (XSEDE) SuperMIC and Stampede2 through the allocation TG-PHY160027, which is supported by National Science Foundation grant number ACI-1548562.
This research also used resources provided by the Los Alamos National
Laboratory Institutional Computing Program, which is supported by the
U.S. Department of Energy National Nuclear Security Administration under
Contract No. 89233218CNA000001. We also used resources provided by
NERSC, which is supported by the US DOE under Contract DE-AC02-05CH11231.

\end{acknowledgments}

\appendix
\section{Equation~(\ref{eq:analytic})}
\label{app}
The two-body problem of three-dimensional scattering with the modified Poschl-Teller potential, Eq.~(\ref{eq:mpt}), can be solved analytically. At low-energies, we have an expression for the phase shift \cite{flu94},
\begin{eqnarray}
\label{eq:ps}
\lim_{q\to 0} \frac{\delta_0}{2q} = \frac{1}{\lambda} - \frac{\pi}{2} \cot \left(
\frac{\pi \lambda}{2	}\right) + \sum_{n=1}^\infty \left( \frac{1}{\lambda+n}-\frac{1}{n} \right) = \nonumber \\
\frac{1}{\lambda} - \frac{\pi}{2} \cot \left(
\frac{\pi \lambda}{2	}\right) + \sum_{n=1}^\infty \frac{-\lambda}{n(\lambda+n)},
\end{eqnarray}
where $q=k/(2\mu)$. We can use the following relations \cite{abr64},
\begin{flalign}
\Psi(1+z)&=-\gamma + \sum_{n=1}^\infty \frac{z}{n(n+z)} \quad (z\neq -1,-2,\cdots), \nonumber \\
\Psi(1+z)&=\Psi(z)+\frac{1}{z},
\end{flalign}
to cast the Eq.~(\ref{eq:ps}) in the form
\begin{eqnarray}
\lim_{q\to 0} \frac{\delta_0}{2q} = - \frac{\pi}{2} \cot \left(
\frac{\pi \lambda}{2	}\right) -\gamma - \Psi(\lambda).
\end{eqnarray}
Approximating $\delta_0 \approx -ka$ yields Eq.~(\ref{eq:analytic}).

\bibliography{article.bib}
\end{document}